\documentclass{article}%
\usepackage{amsfonts}
\usepackage{amsmath}
\usepackage{amssymb}
\usepackage{graphicx}%
\begin{document}

\title{Estimations of total mass and\\energy of the universe}
\author{Dimitar Valev\\\textit{Stara Zagora Department, Solar-Terrestrial Influences Laboratory,}\\\textit{Bulgarian Academy of Sciences, 6000 Stara Zagora, Bulgaria}}
\maketitle

\begin{abstract}
The recent astronomical observations indicate that the expanding universe is
homogeneous, isotropic and asymptotically flat. The Euclidean geometry of the
universe enables to determine the total gravitational and kinetic energy of
the universe by Newtonian gravity in a flat space. By dimensional analysis, we
have found the mass of the universe close to the Hoyle-Carvalho formula $M\sim
c^{3}/(GH)$. This value is independent from the cosmological model and infers
a size (radius) of the universe close to Hubble distance. It has been shown
that almost the entire kinetic energy of the universe ensues from the
cosmological expansion. Both, the total gravitational and kinetic energies of
the universe have been determined in relation to an observer at an arbitrary
location. The relativistic calculations for total kinetic energy have been
made and the dark energy has been excluded from calculations. The total
mechanical energy of the universe has been found close to zero, which is a
remarkable result. This result supports the conjecture that the gravitational
energy of the universe is approximately balanced with its kinetic energy of
the expansion.


\end{abstract}

\section{Introduction}

The problem for the average density of the universe $\overline{\rho}$ acquires
significance when it has been shown that the General Relativity allows to
reveal the geometry and evolution of the universe by simple cosmological
models \cite{Friedmann, Lemaitre, Einstein}. Crucial for the universe appears
dimensionless total density $\Omega=\overline{\rho}/\rho_{c}$, where $\rho
_{c}$ is the critical density of the universe. In the case of $%
\Omega
<1$ (open universe) the global spatial curvature is negative and the geometry
of the universe is hyperbolic and in the case of $%
\Omega
>1$ (closed universe) the curvature is positive and the geometry is spherical.
In the special case of $%
\Omega
=1$ (flat universe) the curvature is zero and the geometry is Euclidean. Until
recently scarce information has been available about the density and geometry
of the universe. The most trustworthy total matter density $\Omega$ has been
determined by measurements of the dependence of the anisotropy of the Cosmic
Microwave Background ($CMB$) upon the angular scale. The recent results show
that $\Omega\approx1\pm\Delta\Omega$, where the error $\Delta\Omega$ decreases
from 0.10 \cite{de Bernardis, Balbi} to 0.02 \cite{Spergel}, i.e. the density
of the universe is close to the critical one and the universe is
asymptotically flat (Euclidean).

The fact that $\Omega$ is so close to a unit is not accidental since only at $%
\Omega
=1$ the geometry of the universe is flat and the flat universe was predicted
from the inflationary theory \cite{Guth}. The total density $\Omega$ includes
matter density $\Omega_{M}=\Omega_{b}+\Omega_{c}$, where $\Omega_{b}%
\approx0.05$ is density of baryon matter and $\Omega_{c}\approx0.22$ is
density of cold dark matter \cite{Peacock}, and dark energy $\Omega_{\Lambda
}\approx0.73$ \cite{Hinshaw} producing an accelerating expansion of the
universe \cite{Riess, Perlmutter}. The found negligible $CMB$ anisotropy
$\delta T/T\sim10^{-5}$ indicates that the early universe was very homogeneous
and isotropic \cite{Bennett}. Three-dimensional maps of the distribution of
galaxies corroborate homogeneous and isotropic universe on large scales
greater than $100$\ $\ Mps$ \cite{Shectman, Stoughton}.

Usually, Einstein pseudotensor is used for determination of the total energy
of the universe \cite{Rosen, Johri}. This approach is general for open, close
and flat anisotropic models, but pseudotensorial calculations are dangerous as
they are very coordinate dependent and thus, they may lead to ambiguous
results \cite{Banerjee}. Newtonian gravity still works reasonably in
practically all gravitational problems, starting from Earth gravity, space
flights and star systems and ending to birth of stars and star clusters, with
exception of extremely compact objects like black holes and neutron stars
possessing strong gravitational field causing non negligible curvature of the
space. After recent $CMB$ observations discovered that the global geometry of
the universe is flat, some cosmological problems could be solved by Newtonian
gravity in Euclidean space. This opportunity has been used in the paper to
estimate total mechanical energy of the universe.

To determine gravitational and kinetic energy of the universe, information of
the size and total mass of the universe are needed. There are different
estimations of the mass of the universe covering very large interval from
$3\times10^{50}%
\operatorname{kg}%
$ \cite{Hopkins} to $1.6\times10^{60}%
\operatorname{kg}%
$ \cite{Nielsen}. Also the estimations of the size (radius) of the universe
are from $10$\ $Glyr$ \cite{Hilgevoord} to more than of $78$\ $Glyr$
\cite{Cornish}. In the paper, we have found the mass of the observable
universe by an original approach for cosmology, namely dimensional analysis.

\section{Estimations of the total mass and size (radius) of the universe}

Taking into account uncertainties of the estimations for the mass and size of
the universe, an original approach for cosmology, namely dimensional analysis,
has been applied below for estimation of the mass and size of the universe.
The dimensional analysis is a conceptual tool often applied in physics to
understand physical situations involving certain physical quantities. It is
routinely used to check the plausibility of derived equations and
computations. When the certain quantity, with which other determinative
quantities would be connected, is known but the form of this connection is
unknown, a dimensional equation was composed for its finding. Most often, the
dimensional analysis is applied in mechanics where there are many problems
having a few determinative quantities. The quantity of mass dimension in high
energy physics is also obtained by means of the fundamental constants $c$, $G$
and $\hbar$. This is the famous Planck mass $m_{P}\sim\sqrt{\hbar c/G}%
\approx2.17\times10^{-8}%
\operatorname{kg}%
$, whose energy equivalent -- the Planck energy $E_{P}=m_{P}c^{2}\sim10^{19}%
\operatorname{GeV}%
$ appears unification energy of the fundamental interactions.

The fundamental parameters as the gravitational constant $G$, speed of the
light $c$ and the Hubble constant $H\approx70\ km\ s^{-1}\ Mps^{-1}$
\cite{Mould} determine the global properties of the universe. Therefore, by
means of these parameters, a mass dimension quantity $m_{x}$ related to the
universe could be constructed:%

\begin{equation}
m_{x}=kc^{\alpha}G^{\beta}H^{\gamma} \label{Eqn1}%
\end{equation}

where $k$ is a dimensionless parameter of the order of magnitude of a unit and
$\alpha,\beta$ and $\gamma$ are unknown exponents which have been found by
dimensional analysis.

Taking into account the dimensions of the quantities in formula (\ref{Eqn1})
we obtain the system of linear equations for unknown exponents:%

\begin{align}
\alpha+3\beta &  =0\nonumber\\
-\alpha-2\beta-\gamma &  =0\label{Eqn2}\\
-\beta &  =1\nonumber
\end{align}

The determinant $\Delta$ of the system is:%

\begin{equation}
\Delta=\left\vert
\begin{array}
[c]{ccc}%
1 & 3 & 0\\
-1 & -2 & -1\\
0 & -1 & 0
\end{array}
\right\vert =-1 \label{Eqn3}%
\end{equation}

The determinant $\Delta\neq0$, therefore the system has a unique solution. We
find this solution by Kramer's formulae:%

\begin{align}
\alpha &  =\frac{\Delta_{1}}{\Delta}=3\nonumber\\
\beta &  =\frac{\Delta_{2}}{\Delta}=-1\label{Eqn4}\\
\gamma &  =\frac{\Delta_{3}}{\Delta}=-1\nonumber
\end{align}

Thus, we find the mass $m_{x}$ related to the universe:%

\begin{equation}
m_{x}\sim\frac{c^{3}}{GH}\sim10^{53}%
\operatorname{kg}
\label{Eqn5}%
\end{equation}

This value hits in the large interval for the mass of the universe mentioned
at the end of Section 1 and coincides with Carvalho formula for the mass of
the universe, deduced by totally different approach \cite{Carvalho}. Thus, the
quantity $m_{x}$, obtained by dimensional analysis by means of the fundamental
parameters $c$, $G$ and $H$, represents acceptable estimation of the mass of
the universe. It is worthy to note that this value is independent from the
cosmological model.

The universe is flat and the total density, including dark matter and dark
energy, is $\overline{\rho}=\Omega\rho_{c}\approx\rho_{c}$, where the critical
density of the universe $\rho_{c}$ \cite{Peebles} determines from:%

\begin{equation}
\rho_{c}=\frac{3H^{2}}{8\pi G}\approx9.5\times10^{-27}kg\ m^{-3} \label{Eqn6}%
\end{equation}

Since the universe is homogeneous and isotropic, it appears 3-dimensional
homogeneous sphere for an observer at arbitrary location. From (\ref{Eqn5})
and (\ref{Eqn6}) we obtain:%

\begin{equation}
\overline{\rho}=\Omega\rho_{c}=\frac{3\Omega H^{2}}{8\pi G}=\frac{m_{x}}%
{V}=\frac{3c^{3}}{4\pi R^{3}GH} \label{Eqn7}%
\end{equation}

From (\ref{Eqn7}) we have estimated the size (radius) of the universe ($R$)
close to the Hubble distance $cH^{-1}$:%

\begin{equation}
R=(2/\Omega)^{1/3}\frac{c}{H}\sim cH^{-1} \label{Eqn8}%
\end{equation}

The result, obtained from (\ref{Eqn8}), shows our universe practically
coincides with Hubble sphere.

The mass of the universe would be deduced more precisely by means of the
recent density of the universe $\overline{\rho}=\Omega\rho_{c}=3\Omega
H^{2}/(8\pi G)$ and radius (size) of the universe $R\sim cH^{-1}$:%

\begin{equation}
M=4\pi R^{3}\overline{\rho}/3\approx\frac{4\pi c^{3}\Omega\rho_{c}}{3H^{3}%
}=\frac{c^{3}\Omega}{2GH} \label{Eqn9}%
\end{equation}

This expression is more accurate than (\ref{Eqn5}), since the results of the
dimensional analysis are correct with accuracy to a coefficient $k\sim1$. This
value practically coincides with the Fred Hoyle formula for the mass of the
universe $M=c^{3}/(2GH)$ \cite{Hoyle}. Any possible matter beyond the Hubble
sphere does not affects the observer. Hence, it has no contribution in the
mass and energy of the universe, calculated in relation to the observer.

Besides, we can estimate the total rest energy of the universe from
(\ref{Eqn9}) and Einstein equation:%

\begin{equation}
E_{0}=Mc^{2}=\frac{c^{5}\Omega}{2GH} \label{Eqn10}%
\end{equation}

\section{Determination of the total mechanical energy of the universe}

The results of dimensional analysis and $CMB$ observations suggest that the
universe appears homogeneous 3-dimensional sphere with radius $R$ close to
Hubble distance $cH^{-1}$. Hence, the gravitational potential energy $U$ of
the universe is:%

\begin{equation}
U=-G%
{\textstyle\int\limits_{0}^{R}}
\frac{M(r)dm}{r}=-\frac{16}{3}G\pi^{2}\overline{\rho}^{2}%
{\textstyle\int\limits_{0}^{R}}
r^{4}dr=-\frac{3GM^{2}}{5R} \label{Eqn11}%
\end{equation}

where 0 is an arbitrary location of the observer, $R\sim cH^{-1}$ is the
radius of the universe and $M(r)=4\pi\overline{\rho}r^{3}/3$ is the mass of a
sphere with radius $r$.

According to the equivalence of mass and energy, dark energy also possesses
mass and gravitational energy. Replacing (\ref{Eqn8}) and (\ref{Eqn9}) in
(\ref{Eqn11}) we have found the total gravitational energy of the universe:%

\begin{equation}
U=-\frac{3c^{5}\Omega^{2}}{20GH} \label{Eqn12}%
\end{equation}

Similar approach has been used for calculation of the total gravitational
energy of a body arising from gravitational interaction of the body with all
masses of the universe \cite{Woodward, Valev a}.

Taking into account formulae (\ref{Eqn10}) and (\ref{Eqn12}) we find
$U=-\frac{3}{10}\Omega Mc^{2},$ i.e. the modulus of the total gravitational
energy of the universe is close to 3/10 of its total rest energy.

The estimation of the total kinetic energy of the universe $T$ is more
complicated as a result of the diversity of movements of masses in the
universe. We suggest that almost all kinetic energy of the universe is a
result of the cosmological expansion since it includes movement of the
enormous masses (galaxies and clusters of galaxies) with average speed of the
order of magnitude of $c/2$. The rotation curves of galaxies show that the
majority of stars move into the galaxies with speed less than $v_{0}%
=3\times10^{5}\ m\ s^{-1}$ \cite{Sofue}. Besides, on rare occasions, the
peculiar (non-cosmological) velocities of galaxies exceed this value
\cite{Strauss}. On the other hand, the speed of medium-distanced galaxies (and
their stars), as a result of the cosmological expansion is of the order of
magnitude of $c/2=1.5\times10^{8}\ m\ s^{-1}$. Obviously, the kinetic energy
of an \textquotedblleft average star\textquotedblright\ in the universe,
ensuing from its peculiar movement, constitutes less than $(2v_{0}/c)^{2}%
\sim4\times10^{-6}$ part of its kinetic energy, ensuing from the cosmological
expansion, therefore, former should be ignored.

Let us estimate the total kinetic energy of the universe in relation to an
observer at arbitrary location. The total kinetic energy of the universe is
the sum of the kinetic energy of all masses $m_{i}$ moving in relation to the
observer with speed $v_{i}$ determined from Hubble law $v_{i}=Hr_{i}$, where
$r_{i}\leq cH^{-1}$ is the distance between the observer and mass $m_{i}$
placed within the Hubble sphere. Newtonian formula for kinetic energy
$T=\frac{1}{2}%
{\textstyle\sum\limits_{i}}
m_{i}v_{i}^{2}$ was used in \cite{Valev b}, but the distant masses recede from
the observer with speeds comparable with the speed of the light $c$.
Therefore, the relativistic formula for kinetic energy is used below:%

\begin{equation}
T=c^{2}%
{\textstyle\sum\limits_{i}}
m_{i}[(1-v_{i}^{2}/c^{2})^{-1/2}-1] \label{Eqn13}%
\end{equation}

Since, for an arbitrary observer, the universe appears a 3-dimensional
homogeneous sphere having radius $R\sim cH^{-1}$, the sum (\ref{Eqn13}) can be
replaced by the integral:%

\begin{equation}
T=c^{2}%
{\textstyle\int}
[(1-v^{2}/c^{2})^{-1/2}-1]dm=4\pi\overline{\rho}c^{2}%
{\textstyle\int\limits_{0}^{R}}
[(1-v^{2}/c^{2})^{-1/2}-1]r^{2}dr \label{Eqn14}%
\end{equation}

Replacing $v\approx v_{r}$ with expression from Hubble law $v_{r}=Hr$,
equation (\ref{Eqn14}) transforms into:%

\begin{equation}
T=4\pi\overline{\rho}c^{2}%
{\textstyle\int\limits_{0}^{R}}
[(1-H^{2}r^{2}/c^{2})^{-1/2}-1]r^{2}dr=4\pi\overline{\rho}c^{2}I-\frac{4}%
{3}\pi\overline{\rho}c^{2}R^{3} \label{Eqn15}%
\end{equation}

where $I=%
{\textstyle\int\limits_{0}^{R}}
(1-H^{2}r^{2}/c^{2})^{-1/2}r^{2}dr.$

The solution of the integral $I$ is given from equation:%

\begin{equation}
I=-\frac{cr}{2H}(c^{2}/H^{2}-r^{2})^{1/2}+\frac{c^{3}}{2H^{3}}\arcsin\frac
{Hr}{c} \label{Eqn16}%
\end{equation}

Considering (\ref{Eqn8}) and replacing low and upper limits of integration we find:%

\begin{equation}
I=\frac{\pi c^{3}}{4H^{3}} \label{Eqn17}%
\end{equation}

Replacing (\ref{Eqn17}) in (\ref{Eqn15}) we obtain:%

\begin{equation}
T=\frac{\pi^{2}\overline{\rho}c^{5}}{H^{3}}-\frac{4\pi\overline{\rho}c^{5}%
}{3H^{3}} \label{Eqn18}%
\end{equation}

In consideration of $\overline{\rho}=\Omega\rho_{c}$ and (\ref{Eqn6}) we find
the total kinetic energy of the universe:%

\begin{equation}
T=\frac{c^{5}\Omega}{2GH}(\frac{3\pi}{4}-1)=Mc^{2}(\frac{3\pi}{4}-1)
\label{Eqn19}%
\end{equation}

Taking into account (\ref{Eqn12}) and (\ref{Eqn19}), the total mechanical
energy of the universe determines from:%

\begin{equation}
E=T+U=\frac{c^{5}\Omega}{2GH}(\frac{3\pi}{4}-1-\frac{3\Omega}{10}%
)=Mc^{2}(\frac{3\pi}{4}-1-\frac{3\Omega}{10})\approx1.056Mc^{2}\sim Mc^{2}
\label{Eqn20}%
\end{equation}

Thus, the total mechanical energy of the universe is found close to its total
rest energy. It is worthy to note that in the process of deducing the formula
(\ref{Eqn19}) the dark energy was accepted as involved in cosmological
expansion ($\Omega=\Omega_{M}+\Omega_{\Lambda}$). But the dark energy has no
kinetic energy, therefore it should be excluded from calculations of the total
kinetic energy of the universe. In result, the density $\overline{\rho}%
=\Omega\rho_{c}$ in formula (\ref{Eqn18}) must be replaced with $\rho
_{M}=\Omega_{M}\rho_{c}$ and taking into account formula (\ref{Eqn6}) we find
a relativistic formula for the total kinetic energy of the universe:%

\begin{equation}
T=\frac{c^{5}\Omega_{M}}{2GH}(\frac{3\pi}{4}-1)=Mc^{2}\Omega_{M}(\frac{3\pi
}{4}-1)\approx1.356\Omega_{M}Mc^{2} \label{Eqn21}%
\end{equation}

The recent value of matter density is between $\Omega_{M}=0.19$
\cite{Carlberg} and $\Omega_{M}=0.27$ \cite{Hinshaw}. As a result, the total
kinetic energy of the universe $T\approx(0.26\div0.37)Mc^{2}$, i.e. close to
3/10 of its total rest energy $Mc^{2}$.

Finally, from (\ref{Eqn12}) and (\ref{Eqn21}) we find the total mechanical
energy of the universe:%

\begin{equation}
E=T+U=\frac{c^{5}}{2GH}[\Omega_{M}(\frac{3\pi}{4}-1)-\frac{3\Omega^{2}}%
{10})=Mc^{2}[\frac{\Omega_{M}}{\Omega}(\frac{3\pi}{4}-1)-\frac{3\Omega}%
{10}]\sim0 \label{Eqn22}%
\end{equation}

It is remarkably, that the total mechanical energy of the universe is close to
zero. This result supports the conjecture that the gravitational energy of the
universe is approximately balanced with its kinetic energy of the expansion
\cite{Lightman}. According to formula (\ref{Eqn22}), the total kinetic energy
of the universe is strictly equal to zero in case of $\Omega_{M}=\frac{3}%
{10}\Omega^{2}/(3\pi/4-1)\approx0.22$. This value should be discussed as a
prediction of the suggested model, which the future more accurate observations
are able to test.

\section{Conclusions}

The recent astronomical observations indicate that the expanding universe is
homogeneous, isotropic and asymptotically flat. The Euclidean geometry of the
universe enables to determine the total kinetic and gravitational energies of
the universe by Newtonian gravity in a flat space.

By an original approach for cosmology, namely dimensional analysis, a mass
dimension quantity of the order of $10^{53}%
\operatorname{kg}%
$, related to the universe, has been found close to Hoyle-Carvalho formula for
the mass of the universe. This value is independent from the cosmological
model and infers the size (radius) of the universe close to Hubble distance
$cH^{-1}$.

Both, the total kinetic and gravitational energies of the universe have been
determined in relation to an observer at arbitrary location. Based on the
simple homogeneous and isotropic model of the flat universe which expands
according to Hubble law, we have found equation for the total gravitational
energy of the universe. The modulus of the total gravitational energy of the
universe is estimated to 3/10 of its total rest energy $Mc^{2}$.

The relativistic calculations for total kinetic energy have been made and the
dark energy has been excluded from calculations. The total kinetic energy of
the universe has been found close to the modulus of its total gravitational
energy. Therefore, the total mechanical energy of the universe is close to
zero, which is a remarkable result. This result supports the conjecture that
the gravitational energy of the universe is approximately balanced with its
kinetic energy of the expansion.

\end{document}